\documentclass[conference]{IEEEtran}
\usepackage[compatibility=false]{caption}
\IEEEoverridecommandlockouts
\usepackage[hidelinks, bookmarks=false]{hyperref}
\usepackage{nomencl}
\usepackage{amsmath}
\usepackage{amssymb}
\usepackage{tabularx}
\usepackage{bbm}
\usepackage[section]{algorithm}
\usepackage{algorithmic}
\usepackage{multirow}
\usepackage{rotating}
\usepackage{pdfpages}
\usepackage{hyphenat}
\usepackage{subfig}
\usepackage{tikz}
\usepackage{tikzpagenodes}
\usepackage{graphicx}
\usepackage{subcaption}
\usepackage[percent]{overpic}
\usepackage{orcidlink}
\usepackage[capitalise]{cleveref} 
\usepackage[toc, acronym]{glossaries}

\glsdisablehyper

\newacronym{snr}{SNR}{signal-to-noise ratio}
\newacronym{osnr}{OSNR}{optical signal-to-noise ratio}
\newacronym{ber}{BER}{bit error rate}
\newacronym{fso}{FSO}{free-space optical}
\newacronym{smf}{SMF}{single-mode fiber}
\newacronym{edfa}{EDFA}{erbium-doped fiber amplifier}
\newacronym{dsp}{DSP}{digital signal processing}
\newacronym{prbs}{PRBS}{pseudo-random binary sequence}
\newacronym{rrc}{RRC}{root raised cosine}
\newacronym{qpsk}{QPSK}{quadrature phase-shift keying}
\newacronym{awg}{AWG}{arbitrary waveform generator}
\newacronym{dpiq}{DP-IQ}{dual-polarization in-phase and quadrature modulator}
\newacronym{ecl}{ECL}{external cavity laser}
\newacronym{aom}{AOM}{acousto-optic modulator}
\newacronym{ase}{ASE}{amplified spontaneous emission}
\newacronym{rf}{RF}{radio frequency}
\newacronym{wss}{WSS}{wavelength selective switch}
\newacronym{dso}{DSO}{digital storage oscilloscope}
\newacronym{osa}{OSA}{optical spectrum analyzer}
\newacronym{cma}{CMA}{constant modulus algorithm}
\newacronym{bps}{BPS}{blind phase search}
\newacronym{acm}{ACM}{adaptive coding modulation}
\newacronym{csi}{CSI}{channel state information}
\newacronym{gbps}{Gbps}{Gigabits per second}
\newacronym{awgn}{AWGN}{additive white gaussian noise}
\newacronym{evm}{EVM}{error vector magnitude}
\newacronym{cfo}{CFO}{carrier frequency offset}

\begin{document}

\title{Blind SNR Estimation for FSO Communication Systems with Deep Fading}

\author{\IEEEauthorblockA{Humberto V. Q. Melo\textsuperscript{*}, Robson A. Colares and Darli A. A. Mello}
\IEEEauthorblockA{\textsuperscript{1}DECOM, Universidade Estadual de Campinas (Unicamp), 400 Albert Einstein Ave., 13083-852, Campinas, SP, Brazil \\
\textit{\textsuperscript{*}h172417@dac.unicamp.br}, \orcidlink{0000-0002-0496-4734}{0000-0002-0496-4734}\\
}
\thanks{This work has been submitted to the IEEE for possible publication. Copyright may be transferred without notice, after which this version may no longer be accessible.}
}

\maketitle

\begin{abstract} 
We evaluate the M2M4 and EVM methods for real-time SNR estimation in FSO communication systems subject to deep fading. Using an experimental setup with controlled deep fading, we show that the M2M4 estimator reliably tracks the SNR profile, making it suitable for triggering transceiver adaptation.


  
\end{abstract}


\section{Introduction}

Over recent years, \gls{fso} communications systems have consolidated as a compelling complement to both \gls{rf} and fiber-based systems, offering high data rates over wireless links using unlicensed spectrum \cite{jahid2022}.
Despite its potential, \gls{fso} communications is severely impaired by atmospheric turbulence, causing deep signal fading \cite{bibi2025}. A natural mitigation strategy is \gls{acm}, whereby the modulation order and code rate are dynamically adjusted in response to the estimated \gls{csi} at the receiver \cite{mouhammad2026}. However, effective \gls{acm} requires accurate and timely estimation of a link quality metric from the received signal \cite{guiomar2024}, such as the receiver \gls{snr} or the pre-foward error correction (FEC) bit error rate (BER).  One possibility is to estimate the pre-FEC BER directly from the error correcting code, but this approach has an intrinsic latency and eventual inaccuracy in severe fading scenarios, particularly for soft-decision schemes. A significant challenge in SNR estimation for digital communications is that it can become unreliable during severe deep fades, potentially leading to ineffective adaptation decisions, such as incorrect modulation-format selection or improper symbol-rate adjustment \cite{guiomar2022}. In addition, in FSO receivers, deep fades may degrade phase recovery, affecting phase-sensitive methods \cite{valjus2025}.

In this paper, we evaluate the M2M4 and error-vector-magnitude-based (EVM-based) \gls{snr} estimators in optical systems with deep fading. Based on the second and fourth moments of the received signal, the M2M4 method has the advantage of providing phase-insensitive SNR estimates. Both methods are assessed using an experimental transmission setup with a time-varying \gls{snr} profile that emulates severe fading conditions.



\section{SNR estimators}
\label{sec:estimators}

\Gls{evm} is widely accepted as a figure of merit to evaluate the quality of communication systems \cite{mahmoud2009}. \gls{evm} can also be used to estimate \gls{snr} as 
\vspace{-0.1cm}
\begin{equation}     \widehat{\mathrm{SNR}}_{\mathrm{EVM}} = \frac{1}{\mathrm{EVM}^2}=\frac{
P_0 
}{
\frac{1}{N}\sum_{n=1}^{N}\left|S_r(n)-S_t(n)\right|^2},
\label{eq:snr_EVM}
\end{equation}
where $S_r(n)$ is the $n\text{th}$ received symbol, $S_t(n)$ is the $n\text{th}$ transmitted symbol,  and $P_0$ is the average power of all symbols for the chosen modulation. In unsupervised algorithms, $S_t(n)$ is estimated blindly at the receiver.

Another well-known \gls{snr} estimator in the literature is the M2M4 estimator. It uses the second and fourth moments of the received signal, estimated as $ M_2 = \frac{1}{N} \sum_{n=0}^{N-1} |S_r(n)|^2$ and $M_4 = \frac{1}{N} \sum_{n=0}^{N-1} |S_r(n)|^4$. For \gls{qpsk} signals, the M2M4 method estimates the \gls{snr} as \cite{pauluzzi2000}


\begin{equation}
    \widehat{\textrm{SNR}}_{M2M4} = \frac{\sqrt{2 M_2^2 - M_4 }}{M_2 - \sqrt{2 M_2^2 - M_4 }}.
    \label{eq:snr_m2m4}
\end{equation}

Both \cref{eq:snr_EVM} and \cref{eq:snr_m2m4} provide accurate SNR estimation  for \gls{awgn} channels, especially for regions of high \gls{snr}. However, their ability to periodically track \gls{snr} in practical scenarios, where $S_r(n)$ is impaired by residual phase noise, \gls{cfo}, or FSO deep fades, still requires a more careful evaluation.



\section{Experimental Setup}
\label{sec:setup}

\begin{figure}[t]
    \centering
    \includegraphics[width=0.9\linewidth]{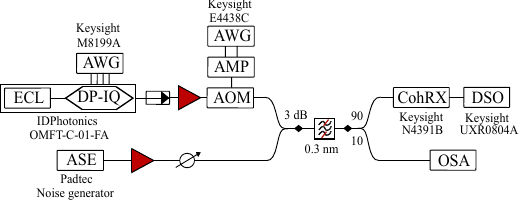}
    \caption{\small{Experimental setup for transmission with a controlled time-varying \gls{snr}.}}
    \vspace{-0.6cm}
    \label{fig:setup}
\end{figure}

 \cref{fig:setup} describes the experimental setup. A $2^{15}-1$ \gls{prbs} is mapped onto a \gls{qpsk} constellation and pulse shaped by a \gls{rrc} filter. The symbol rate is set to $Rs=4~\mathrm{Gbaud}$. The electrical waveform generated by the \gls{awg}, Keysight M8199A, drives the \gls{dpiq}, which modulates the optical carrier provided by an \gls{ecl}. The modulated optical signal is amplified and forwarded to an \gls{aom} that imprints the desired attenuation profile. The \gls{aom} is driven by an a electrical waveform generated by a second \gls{awg}. \Gls{ase} noise is generated by a noise source and amplified by an \gls{edfa}. The modulated optical signal and the \gls{ase} component are combined using a 3-$\mathrm{dB}$ coupler. The resulting signal is then filtered by a \gls{wss} to suppress out-of-band \gls{ase} noise. Finally, the data is received and stored by an optical modulation analyzer.

We use the \gls{osa} to measure \gls{snr} and set a ceiling for the \gls{snr} varying profile. To run the experiment, the \gls{awg} sends a triangular electrical signal with 250 $\mu s$ period and modulated at $200~\mathrm{MHz}$ to drive the \gls{aom} and impose a periodic attenuation profile to the optical signal power while maintaining the noise power constant. 

\section{Results}
\label{sec:results}





We process the data stored by the DSO offline. We use a \gls{dsp} chain composed by coarse frequency recovery,  \gls{cma} equalization, fine frequency recovery and the \gls{bps} algorithm for phase recovery. In systems employing \gls{bps}, the presence of high additive noise induces the occurrence of cycle slips leading to catastrophic error bursts. Therefore, to circumvent this problem, we use differential encoding and decoding. Also, the presence of deep fades requires a resync process if synchronization is lost. In this paper, we use \gls{snr} estimates as a metric to indicate the necessity of resync. We resync the recovered bit sequence with the transmitted \gls{prbs} whenever the estimated \gls{snr} crosses a $0$~dB threshold with a positive derivative.

\begin{figure}[t]
    \centering
    \begin{minipage}{\linewidth}
        \centering
        \begin{overpic}[width=\linewidth]{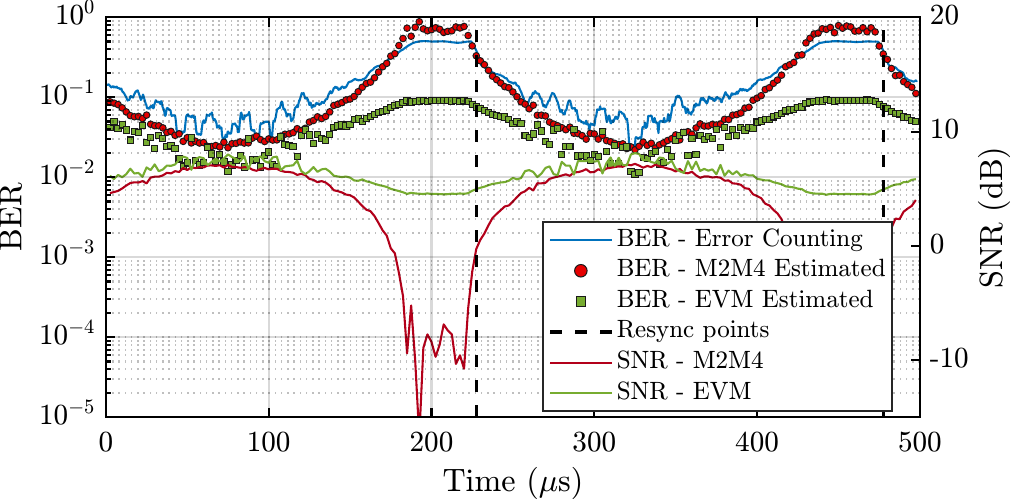}
            \put(0,48){\footnotesize (a)} 
        \end{overpic}
    \end{minipage}

    \vspace{1em} 

    \begin{minipage}{\linewidth}
        \centering
        \begin{overpic}[width=\linewidth]{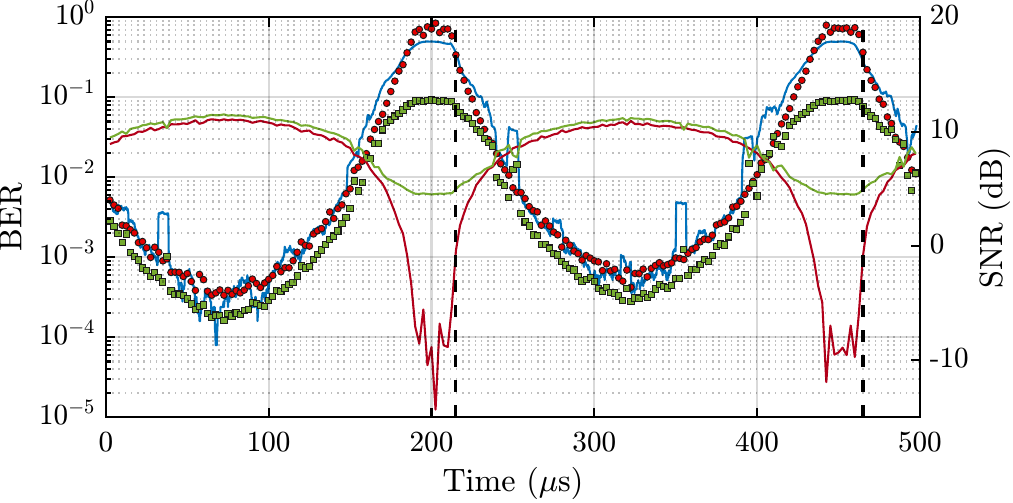}
            \put(0,48){\footnotesize (b)} 
        \end{overpic}
    \end{minipage}

    \caption{\small{\gls{ber} and estimated \gls{snr} evolution over time for a \gls{snr} ceiling of (a) 7 dB and (b) 10 dB, comparing the M2M4 and \gls{evm} estimated \gls{ber} against error counting \gls{ber}. The resync points triggered by the \gls{snr} threshold are indicated with vertical lines.}}
    \vspace{-0.5cm}
    \label{fig:BER_SNR_over_time}
\end{figure}

After the \gls{dsp} chain and the required resync, we compute the \gls{ber} by error counting and compare it with the \gls{ber} derived from SNR values estimated by M2M4 and \gls{evm}, evaluating the ability of the investigated methods to estimate SNR values that are compatible with the true pre-FEC BER. \cref{fig:BER_SNR_over_time} shows the SNR and \gls{ber} evolution over time for two \gls{snr} ceiling scenarios, comparing the M2M4 and \gls{evm} derived \gls{ber} against the \gls{ber} computed by error counting. The \gls{snr} is estimated using \cref{eq:snr_m2m4} and \cref{eq:snr_EVM} in a block-wise manner, with blocks of $10000$ symbols, corresponding to an estimation period of $ 3~\mu \mathrm{s}$. The SNR-derived BER is computed using $\widehat{\mathrm{BER}} = \operatorname{erfc}\!\left(\sqrt{{\mathrm{\widehat{SNR}/2 }}}\right)$ (accounting for differential encoding), while the error counting \gls{ber} curve is computed by averaging in a moving window of 50,000 bits. The resync points are indicated by the vertical dashed lines and the estimated \gls{snr} profile is plotted as a reference.

 \cref{fig:BER_SNR_over_time}(a) presents the case  with an \gls{snr} ceiling of approximately 
$7\,\mathrm{dB}$. The M2M4 estimated \gls{ber} closely follows the error counting curve, even at higher \gls{ber} values, while the 
\gls{evm} estimator exhibits larger deviations in this regime. 
\cref{fig:BER_SNR_over_time}(b) increases the \gls{snr} ceiling to 
$10\,\mathrm{dB}$, revealing an improved agreement between both estimators. Still, the M2M4 
consistently provides a superior BER tracking performance across the 
full \gls{ber} range. This behavior is explained by the reduced impact of 
residual phase errors at higher \gls{snr}, enhancing the performance of the EVM-based method. These results demonstrate that M2M4 is a more reliable 
estimator of SNR and pre-FEC \gls{ber} for optical systems with time-varying \gls{snr} 
profiles, and could be used to trigger adaptive coding and modulation events in the presence of deep fading.

\section{Conclusion}
\label{sec:conclusion}

We evaluate the M2M4 and \gls{evm} \gls{snr} estimators for signal quality estimation in optical systems with a time-varying \gls{snr} profile. Using an experimental transmission setup with a controlled periodic fading profile, we evaluate the estimated SNR and \gls{ber} for two \gls{snr} ceiling scenarios. The results show that the BER derived from the M2M4 estimator closely tracks the error counting \gls{ber} curve across the full \gls{ber} range, outperforming the \gls{evm} estimator particularly at lower \gls{snr} values. The agreement improves at higher \gls{snr} ceilings, as the impact of eventual residual phase errors is reduced. These findings demonstrate that the M2M4 method provides reliable real-time quality metrics for triggering \gls{acm} in optical systems subject to deep fading.

\end{document}